\documentclass[useAMS,usenatbib]{mn2e}

\usepackage{graphicx}

\newcommand\rs[1]{_\mathrm{#1}}
\def\gsim{\;\lower4pt\hbox{${\buildrel\displaystyle >\over\sim}$}\,}

\title[Modeling the blast wave of the nova V407~Cygni]{Modeling the 2010
blast wave of the symbiotic-like nova V407~Cygni}

\author[S. Orlando, J.J. Drake]{Salvatore Orlando$^{1}$
\thanks{E-mail: orlando@astropa.inaf.it},
 Jeremy J. Drake$^{2}$\\ \\
$^{1}$INAF - Osservatorio Astronomico di Palermo ``G.S. Vaiana'',
       Piazza del Parlamento 1, 90134, Palermo, Italy\\
$^{2}$Harvard-Smithsonian Center for Astrophysics, 60 Garden
              Street, Cambridge, MA 02138, USA
}
\begin{document}

\date{Accepted. Received}

\pagerange{\pageref{firstpage}--\pageref{lastpage}} \pubyear{}

\maketitle

\label{firstpage}

\begin{abstract}
The symbiotic-like binary Mira and nova V407~Cyg was observed
in outburst on March 2010 and monitored in several wavelength bands. The
outburst had, to some extent, characteristics similar to those observed
during other nova eruptions, such as recently occurred in RS~Oph and
U~Sco, suggesting that the blast wave interacted with the giant companion
and propagated through a dense circumstellar medium enveloping the binary
system. Here we report on multi-dimensional hydrodynamic simulations
describing the 2010 outburst of V407~Cyg, exploring the first 60 days of
evolution. The model takes into account  thermal conduction (including the
effects of heat flux saturation) and radiative cooling; the pre-explosion
system conditions included the companion star and a circumbinary density
enhancement that are believed to influence the evolution and morphology
of the blast wave. The simulations showed that the blast and the ejecta
distribution are both aspherical due to the inhomogeneous circumstellar
medium in which they expand; in particular they are significantly
collimated in polar directions (producing a bipolar shock morphology)
if the circumstellar envelope is characterized by an equatorial density
enhancement. The blast is partially shielded by the Mira companion,
producing a wake with dense and hot post-shock plasma on the rear side
of the companion star; most of the X-ray emission produced during
the evolution of the blast arises from this plasma structure. 
The observed X-ray lightcurve can be reproduced, assuming values
of outburst energy and ejected mass similar to those of RS~Oph and
U~Sco, if a circumbinary gas density enhancement is included in the
model. In particular, our ``best-fit'' model predicts that the 2010 blast
propagated through a circumbinary gas density enhancement with radius
of the order of 40 AU and gas density $\approx 10^6$~cm$^{-3}$ and
that the mass of ejecta in the outburst was $M\rs{ej} \approx 2\times
10^{-7}~M\rs{\odot}$ with an explosion energy $E\rs{0} \approx 2\times
10^{44}$~erg. Alternatively, the model can produce a similar X-ray
lightcurve without the need of a circumbinary gas density enhancement
only if the outburst energy and ejected mass were similar to those at
the upper end of ranges for classical novae, namely $M\rs{ej} \approx
5\times 10^{-5}~M\rs{\odot}$ and $E\rs{0} \approx 5\times 10^{46}$~erg.
\end{abstract}


\section{Introduction}
\label{s:intro}

The symbiotic binary V407~Cyg was discovered in outburst at 7th
magnitude by \citet{2010IAUC.9130....1N}  on 2010 March 10, in stark
contrast to occasional brightenings observed over the years by
a few magnitudes compared to historical low states in the range
13--$17 m\rs{pg}$.  The binary comprises a white dwarf (WD) and
Mira-type M6-7 red giant pulsating with a period of 763d, with an
orbital period of 43~yr \citep{1990MNRAS.242..653M,2003ARep...47..777K}.
Spectroscopy obtained by \citet{2011MNRAS.410L..52M} and
C.~Buil\footnote{http://www.astrosurf.com/~buil/v407cyg/obs.htm} two days
after outburst showed very broad Balmer lines; \citet{2011MNRAS.410L..52M}
reported a full-width at half-maximum (FWHM) of 2760~km~s$^{-1}$ on
day 2.3,  and described the outburst as a He/N nova expanding within
the wind of the Mira companion, similar to the 2006 explosion of
RS~Oph \citep[see][and references therein]{2008ASPC..401.....E}.   SiO maser
emission was found to decline dramatically after the burst, leading
\citet{2011PASJ...63..309D} to conclude that the maser emitting regions were
essentially wiped out in a time scale of two weeks by the propagation
of the nova shock into the Mira envelope.

The RS~Oph blast wave produced copious X-rays from at most 3
days after optical discovery, when it was first observed by {\it
Swift} and {\it RXTE}, characterized by a steady decline thereafter
\citep{2006Natur.442..276S, 2006ApJ...652..629B, 2008ApJ...673.1067N,
2009ApJ...691..418D}.  V407~Cyg also developed into an X-ray source,
though somewhat more slowly, exhibiting a strong brightening at 20 days or
so, a peak near 30 days and a steady decline. \cite{2011A&A...527A..98S}
analyzed the {\it Swift} UV and X-ray lightcurves together with
optical spectra obtained during the first three months of the
eruption.  Balmer lines showed steady secular narrowing \citep[see
also][]{2011MNRAS.410L..52M}, and lines of [Ca~V], [Fe~VII], [Fe~X],
and He~II exhibited asymmetric profiles attributed to an aspherical
expansion reminiscent of that diagnosed for RS~Oph from X-ray lines by
\citet{2009ApJ...691..418D}.

While not nearly as bright as RS~Oph in X-rays, V407~Cyg was in some ways
even more remarkable, with a firm detection by the {\em Fermi} Large
Area Telescope of variable $\gamma$-ray emission in the 0.1--10~GeV
range on 2010~March~10---the same day as optical discovery---that
persisted for two weeks \citep{2010Sci...329..817A}.  This was the first
ever $\gamma$-ray detection of a nova explosion and was attributed
by \citet{2010Sci...329..817A} to $\pi^0$ production and subsequent
decay resulting from collisions of protons accelerated in the shock.
\citet{2011MNRAS.413L..11L} pointed out that short period symbiotic novae
such as RS~Oph are unlikely to produce $\gamma$-rays because of the fast
evolution of the blast wave in the higher density environment closer in to
the red giant, whereas long period systems such as V407~Cyg provide a much
longer-lived accelerator.  \citet{2010PhRvD..82l3012R} further noted that
V407~Cyg  could have been a detectable source of high energy neutrinos.

The outburst mechanism for novae is thermonuclear runaway on
the WD triggered by the mass of accreted material exceeding
a critical limit \citep{1985ApJ...293L..23S, 1988ApJ...325L..35S}.
In two earlier papers we have applied sophisticated hydrodynamic
models to the nova explosions on RS~Oph \citep{2009A&A...493.1049O}
and U~Sco \citep{2010ApJ...720L.195D} to provide key insights into the
nature of the explosion and its environment. In both sets of models,
circumbinary material proved crucial in explaining the observed X-ray
emission while providing a degree of collimation to the explosions. Here
we describe similar detailed hydrodynamic simulations of the V407~Cyg
explosion, and pay particular attention to the effects of circumbinary
gas density enhancement and the secondary Mira companion on the explosion.
In Sect.~\ref{s:obsanal} we describe the hydrodynamic model, the numerical
setup, and the synthesis of X-ray emission; in Sect. \ref{s:discuss}
we discuss the results; and finally in Sect. \ref{s:conclusion} we draw
our conclusions.

\section{Hydrodynamic Modeling}
\label{s:obsanal}

The blast wave was modeled by numerically solving the time-dependent
fluid equations of mass, momentum, and energy conservation,
including radiative losses described by an optically-thin plasma
and thermal conduction; the latter incorporated the effects of heat flux
saturation. Owing to the long timescale of the blast evolution,
radiative losses and thermal conduction were found to be more important
than in the cases of RS~Oph (\citealt{2009A&A...493.1049O}) and U~Sco
(\citealt{2010ApJ...720L.195D}). The long evolution timescale of the blast
coupled with large expansion velocities of a few thousand km~s$^{-1}$
rendered the spatial extent too large and computationally demanding
to perform an extensive set of fully 3-dimensional (3D) hydrodynamic
simulations while still resolving the structure of the immediate binary
environment, even with many levels of adaptive mesh refinement. The
computational cost is raised by the inclusion of thermal conduction that
is solved explicitly, such a scheme being subject to a rather restrictive
stability condition as the thermal conduction timescale is generally
shorter than the dynamical one (e.g. \citealt{2000A&A...362L..41H,
2005CoPhC.168....1H, 2008ApJ...678..274O, 2010A&A...510A..71O}).

Given the large computational cost required by 3D simulations, we
adopted the following strategy: we first explored the wide parameter
space of the model by adopting a 2-dimensional (2D) cylindrical
coordinate system $(r,z)$ and, therefore, assuming the system to
be symmetrical with respect to the axis passing through the WD and
the companion star; then, for the set of parameters found to best
reproduce the observations, we relaxed the hypothesis of axisymmetry
and performed a fully 3D simulation in cartesian geometry. The 2D
and 3D models developed here are otherwise similar to the 3D models
of \citet{2009A&A...493.1049O} and \cite{2010ApJ...720L.195D} and
we refer to those works for further details. The calculations were
performed using FLASH, an adaptive mesh refinement multiphysics code for
astrophysical plasmas \citep{Fryxell2000ApJS} extended with additional
computational modules to handle radiative losses and thermal conduction
(see \citealt{2005A&A...444..505O} for the details of the implementation).
The hydrodynamic equations for compressible gas dynamics are solved
using the FLASH implementation of the piecewice-parabolic method (PPM;
\citealt{1984JCoPh..54..174C}).

We adopted the system parameters of \citet{1990MNRAS.242..653M}; these
are listed in Table~\ref{t:params}. For the hydrodynamic models, the
most important system parameters are the orbit semi-major axis, the
wind mass loss rate and wind terminal velocity. These determine the gas
density into which the blast occurs, and for a given explosion energy
largely control the subsequent evolution timescale of the resulting shock
wave system. The mass loss rate and terminal velocity are based on the
data of \citet{1985ApJ...292..640K}, while the orbital separation was
estimated based on dust extinction changes assumed connected with orbital
modulation. \citet{1990MNRAS.242..653M} estimated orbital separations of
14.0, 15.5 and 16.4~AU for WD masses $M\rs{WD}=0.5$, 1.0 and 1.4~$M_\odot$,
respectively; we adopt 15.5~AU and $M\rs{WD}=1.0\,M_\odot$. We also adopted
the \citet{1990MNRAS.242..653M} distance of 2.7~kpc, derived from the
\citet{1982MNRAS.199..245G} luminosity vs.\ pulsation period relation
extrapolated to the V407~Cyg period of 745 days.

\begin{table*}
\caption{Adopted parameters and initial conditions for the hydrodynamic
models of the 2010 V407~Cyg explosion
\label{t:params}}
\begin{center}
\begin{tabular}{lllllllll}
\hline
\multicolumn{2}{l}{Parameter} & \multicolumn{6}{c}{Value} \\\hline
\multicolumn{2}{l}{Secondary star radius} & \multicolumn{6}{c}{$R\rs{cs}=2.2$ AU} \\
\multicolumn{2}{l}{Binary separation}     & \multicolumn{6}{c}{$a= 15.5$ AU} \\
\multicolumn{2}{l}{Orbital period}        & \multicolumn{6}{c}{$P = 43$ yr} \\
\multicolumn{2}{l}{Spatial domain}        & \multicolumn{3}{c}{(2D cylindrical)} & \multicolumn{3}{c}{(3D cartesian)} \\
     &  & \multicolumn{3}{c}{\hspace{0.53cm}$0 \leq r \leq 530$ AU} & \multicolumn{3}{c}{$-530 \leq x \leq 530$ AU} \\
     &  & \multicolumn{3}{c}{$-530 \leq z \leq 530$ AU} & \multicolumn{3}{c}{$\hspace{0.53cm}0 \leq y \leq 530$ AU} \\
     &  & & & & \multicolumn{3}{c}{$\hspace{0.53cm}0 \leq z \leq 530$ AU} \\
\multicolumn{2}{l}{AMR max. resolution}   & \multicolumn{6}{c}{$5\times
10^{11}$ cm ($3.33\times 10^{-2}$ AU)} \\
\multicolumn{2}{l}{Time covered} & \multicolumn{6}{c}{0--60 days} \\ \hline
            &  & $E_0$ & $M\rs{ej}$ & $n\rs{w}$ & $n\rs{cde}$ & $l_1$ & $l_2$ & $l_3$\\
Model abbreviation & geometry & [erg] & $[M_\odot]$ & [cm$^{-3}$] &   [cm$^{-3}$]  & [AU] & [AU] & [AU] \\
\hline
E44.3-NW7 & 2D cylindrical & $2\times 10^{44}$ & $2\times 10^{-7}$ & $10^7$ & $-$ & $-$ & $-$ & $-$ \\ 
E44.3-NW8 & 2D cylindrical & $2\times 10^{44}$ & $2\times 10^{-7}$ & $10^8$ & $-$ & $-$ & $-$ & $-$ \\
E44.3-NW9 & 2D cylindrical & $2\times 10^{44}$ & $2\times 10^{-7}$ & $10^9$ & $-$ & $-$ & $-$ & $-$ \\
E45-NW8   & 2D cylindrical & $10^{45}$         & $10^{-6}$         & $10^8$ & $-$ & $-$ & $-$ & $-$ \\
E45-NW9   & 2D cylindrical & $10^{45}$         & $10^{-6}$         & $10^9$ & $-$ & $-$ & $-$ & $-$ \\
E46-NW9   & 2D cylindrical & $10^{46}$         & $10^{-5}$         & $10^9$ & $-$ & $-$ & $-$ & $-$ \\
E46.6-NW9 & 2D cylindrical & $5\times 10^{46}$ & $5\times 10^{-5}$ & $10^9$ & $-$ & $-$ & $-$ & $-$ \\
E47-NW10  & 2D cylindrical & $10^{47}$         & $10^{-4}$         & $10^{10}$ & $-$ & $-$ & $-$ & $-$ \\
E44-NW7-CDE6-L40 & 2D cylindrical & $10^{44}$ & $10^{-7}$ & $10^7$ & $10^6$ & 40 & 40 & $-$ \\
E44-NW7-CDE6.7-L40 & 2D cylindrical & $10^{44}$ & $10^{-7}$ & $10^7$ & $5\times 10^6$ & 40 & 40 & $-$ \\
E44-NW7-CDE7-L40 & 2D cylindrical & $10^{44}$ & $10^{-7}$ & $10^7$ & $10^7$ & 40 & 40 & $-$ \\
E44.3-NW7-CDE6.3-L20 & 2D cylindrical & $2\times 10^{44}$ & $2\times 10^{-7}$ & $10^7$ & $2\times 10^6$ & 20 & 20 & $-$ \\
E44.3-NW7-CDE6.3-L40 & 2D cylindrical & $2\times 10^{44}$ & $2\times 10^{-7}$ & $10^7$ & $2\times 10^6$ & 40 & 40 & $-$ \\
E44.3-NW7-CDE6.3-L80 & 2D cylindrical & $2\times 10^{44}$ & $2\times 10^{-7}$ & $10^7$ & $2\times 10^6$ & 80 & 80 & $-$ \\
E44.7-NW7-CDE6.7-L40 & 2D cylindrical & $5\times 10^{44}$ & $5\times 10^{-7}$ & $10^7$ & $5\times 10^6$ & 40 & 40 & $-$ \\ 
E45-NW7-CDE6-L40 & 2D cylindrical & $10^{45}$ & $10^{-6}$ & $10^7$ & $10^6$ & 40 & 40 & $-$ \\
E45-NW7-CDE6.7-L40 & 2D cylindrical & $10^{45}$ & $10^{-6}$ & $10^7$ & $5\times 10^6$ & 40 & 40 & $-$ \\
E45-NW7-CDE7-L40 & 2D cylindrical & $10^{45}$ & $10^{-6}$ & $10^7$ & $10^7$ & 40 & 40 & $-$ \\
3D-E44.3-NW7-CDE6.3 & 3D cartesian & $2\times 10^{44}$ & $2\times 10^{-7}$ & $10^7$ &
$2\times 10^6$ & 53 & 53 & 27 \\ \hline
\end{tabular}
\end{center}
\end{table*}%

\begin{figure}
  \centering
  \includegraphics[width=8truecm]{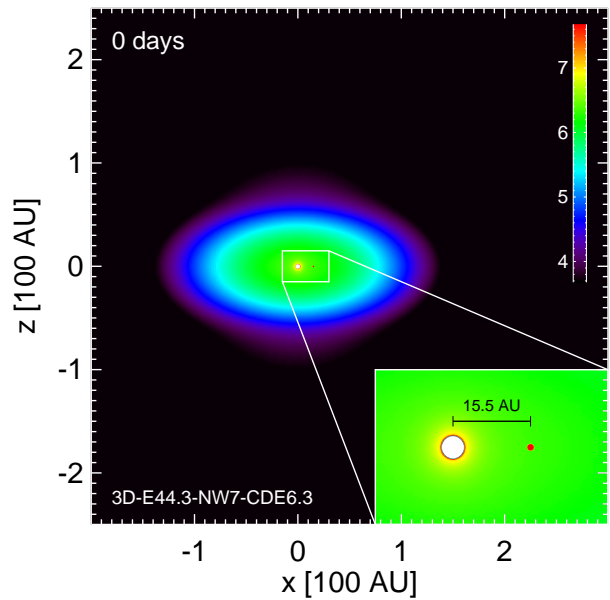}
  \caption{Colour-coded cross-section image of the gas density
  distribution, on a logarithmic scale, in units of cm$^{-3}$, showing
  the initial conditions of run 3D-E44.3-NW7-CDE6.3. The inset panel
  shows a close-up view of the initial geometry of the V407~Cyg
  system. The companion star is at the origin (white circle on the left
  of the inset), and the white dwarf lies on the $x$ axis at $x=15.5$
  AU (red circle on the right).} \label{fig1}
\end{figure}

\begin{figure}
  \centering
  \includegraphics[width=8truecm]{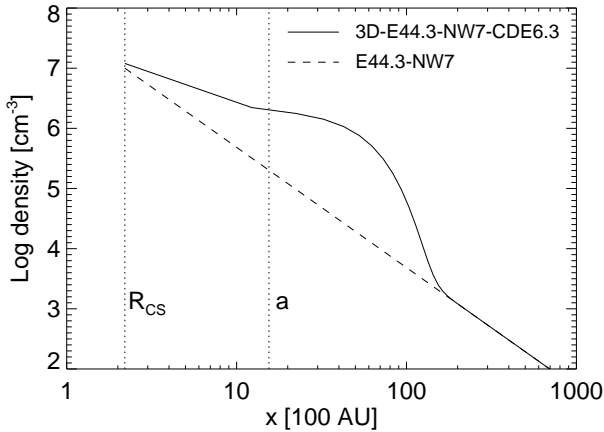}
  \caption{Initial density profiles along the $x$ axis for a model
   including the CDE (run 3D-E44.3-NW7-CDE6.3; solid line) and for a
   model without CDE (run E44.3-NW7; dashed line). The dotted lines
   mark the radius of the Mira companion, $R\rs{cs}$, and the binary
   separation, $a$.}
  \label{fig2}
\end{figure}

Figure~\ref{fig1} shows an example of the initial conditions adopted in
our simulations. The thermonuclear explosion is initiated by a spherical
Sedov-type blast wave \citep{1959sdmm.book.....S} centered on the WD,
with radius $r_{\rm 0} = 10^8$~km (red circle on the right in the inset
panel of Fig.~\ref{fig1}). In analogy with supernova explosions,
the total energy of the blast, $E_{\rm 0}$, is partitioned so that most
of the explosion energy is kinetic (in particular we assumed 1/4 of the
energy contained in thermal energy and the other 3/4 in kinetic energy
as representative; e.g.  \citealt{1996ApJ...471..279D}). The total mass
of the ejecta is $M\rs{ej}$. The blast propagates through the
extended outer atmosphere (the wind) of the Mira companion and is off-set
from the origin of the wind density distribution by 15.5 AU (i.e. the
system orbital separation; see inset panel of Fig.~\ref{fig1}). We
assumed the gas density in the wind is proportional to $R^{-2}$ (where
$R$ is the radial distance from the Mira). In addition to the $R^{-2}$
density distribution, we also included a circumstellar envelope (or a
circumbinary density enhancement in the wind, hereafter CDE), as predicted
by detailed hydrodynamic modeling (e.g.\ \citealt{1999ApJ...523..357M,
2008A&A...484L...9W}). In particular, in systems comprising a
red giant star, these models predict a CDE created by gravitational
accumulation of the cool red giant wind toward the WD and a spiral
shock wave caused by the motion of the stars through this gas.
This scenario is also supported by observations in different
wavelength bands of similar nova outbursts as, for instance,
RS~Oph (e.g. \citealt{2006Natur.442..279O, 2007ApJ...665L..63B,
2008ApJ...688..559R, 2009ApJ...691..418D}). In cartesian geometry,
the mass density distribution of the unperturbed circumstellar medium
(CSM) is given by:

\begin{equation}
\rho = \rho\rs{w}\left(\frac{R\rs{cs}}{R}\right)^2+
\rho\rs{cde}~e^{[-(x/l_1)^2-(y/l_2)^2-(z/l_3)^2]}
\label{csm}
\end{equation}

\noindent
where $\rho\rs{w} = \mu m\rs{H} n\rs{w}$ is the wind mass density close
to the surface of the Mira companion, $\mu = 1.3$ is the mean atomic mass
(assuming cosmic abundances), $m\rs{H}$ is the mass of the hydrogen atom,
$R\rs{cs}$ is the radius of the Mira, $R$ is the radial distance from the
Mira, $\rho\rs{cde} = \mu m\rs{H} n\rs{cde}$ is the density enhancement
close to the Mira, and $l_1$, $l_2$ and $l_3$ are characteristic
length scales determining the size and shape of the CDE. As examples,
Fig.~\ref{fig2} shows the density profiles along the $x$ axis for a model
including the CDE (run 3D-E44.3-NW7-CDE6.3) and for a model without a CDE
(run E44.3-NW7). The Mira companion is included in the calculation as an
impenetrable body with radius $R\rs{cs} = 2.2$ AU (i.e. $\approx 500\,
R_\odot$; white circle on the left in the inset panel of Fig.~\ref{fig1}).

In the 3D simulation, the hydrodynamic equations were solved in one
quadrant of the whole spatial domain with the coordinate system oriented
in such a way that both the WD and the companion star lie on the
$x$ axis (see Fig.~\ref{fig1}; see also \citealt{2010ApJ...720L.195D} for
more details). The companion is at the origin of the coordinate system,
$(x,y,z) = (0,0,0)$, and the computational domain extends 1060~AU in
the $x$ direction and 530~AU in both the $y$ and $z$ directions; the
WD is located to the right on the $x$ axis ($y=z=0$) at $x=15.5$~AU. We
imposed reflecting boundary conditions at $y\rs{min}=0$ and $z\rs{min}=0$
(consistent with the adopted symmetry) and outflow (zero-gradient)
conditions at the other boundaries.

Our 2D simulations considered a slab corresponding to the $(x,z)$ plane
of the 3D simulation, and the hydrodynamic equations were solved in one
half of the 2D spatial domain, taking advantage of the symmetry of the
configuration. In particular, assuming axisymmetry about the axis passing
through the companion star and the WD, the cylindrical coordinate system
was oriented in such a way that both the donor star and the WD lie on
the $z$ axis, with the former at the origin of the coordinate system,
$(r,z) = (0,0)$, and the latter at $z = 15.5$ AU.  Since both
the donor star and the initial nova remnant lie on the symmetry axis
they are modelled as spheres in the cylindrical coordinate system.
The computational domain extended 530 AU in the $r$ direction, and 1060
AU in the $z$ direction. The adopted geometrical system configuration
implied imposition of an axisymmetric (reflecting) boundary condition
at $r\rs{min} = 0$ and outflow conditions at the remaining boundaries.

In order to compare the results of the 2D simulations with those derived
from the 3D simulation, we reconstructed the 3D spatial distributions of
all the physical variables (e.g. mass density and temperature) derived
with the 2D simulations, by rotating the 2D slab about the
$z$ axis, according to the symmetry of the problem. Thus we passed from
the cylindrical coordinate system to the cartesian one, orienting the
cartesian coordinate system in such a way that both the WD and
the companion star lie on the $x$ axis, as in the 3D simulation.

The explosion and subsequent blast wave was followed for a total
of 60 days in order to explore the evolution of the X-ray emission
and study the effects of the circumstellar environment on the
evolution of the blast during the evolutionary phase that was
characterized by a pronounced peak of X-ray emission in the 2010 outburst
(\citealt{2011A&A...527A..98S}). As with the modeling of the U~Sco blast,
the small scale of the stellar system compared with the size of the
rapidly expanding blast wave over the 60 day period of interest presents
a major computational challenge. To this end, we exploited the adaptive
mesh capabilities of the FLASH code by using 12 nested levels of adaptive
mesh refinement, with resolution increasing twice at each refinement
level. This grid configuration yielded an effective maximum resolution of
$\approx 5\times 10^6$ km at the finest level, corresponding to $\approx
20$ grid points per initial radius of the blast and $\approx 66$ grid
points per radius of the Mira companion. The calculations were performed
using an automatic mesh derefinement scheme in the whole spatial domain
except in the Mira (where the resolution does not vary during the blast
evolution) that kept the computational cost approximately constant as the
blast expanded. The effective mesh size was $16384\times 32768$ for the
2D simulations, and $32768\times 16384\times 16384$ for the 3D simulation.

Solar abundances of \citet{1998SSRv...85..161G} (GS) were assumed
for the wind and circumstellar density enhancement, while ejecta
metal abundances were assumed to be enhanced by a factor of
ten. This latter choice was guided by the {\it Swift} spectra that
\citet{2011A&A...527A..98S} found to be rich in both N and O---by
possibly more than a factor of ten---and by observations of He-rich
ejecta in the outburst of U~Sco \citep[see][]{2010ApJ...720L.195D}, as well
as the \citet{2009ApJ...691..418D} high-resolution X-ray spectroscopic
study of the 2006 RS~Oph blast that found evidence for metal-rich
ejecta. The adopted abundances are relevant for the radiative losses
from shocked ejecta, and for the local absorption by the shocked CSM
(with GS abundances) and by the ejecta (with GS abundances $\times 10$)
encountered within the blast wave. The choice of abundances is also
relevant for computation of the emitted X-ray intensity of the blast. Such
emission was synthesized from the model results using the methodology
described by \citet{2009A&A...493.1049O}. The synthesis includes: thermal
broadening of emission lines, Doppler shift of lines due to the component
of plasma velocity along the line-of-sight, photoelectric absorption by
the interstellar medium (ISM), CSM, and ejecta. The absorption by the ISM
was calculated assuming a column density $N_{\rm H} = 2\times 10^{21}$
cm$^{-2}$; the local absorption was calculated self-consistently from
the distributions of shocked CSM and shocked ejecta.

The influence of the different system parameters was investigated
through the 2D simulations, by exploring models with an initial
energy of explosion, $E_0$, in the range $10^{44}$--$10^{47}$~erg,
ejecta mass, $M\rs{ej}$, in the range $10^{-7}$--$10^{-4}$~$M_\odot$,
wind density in the range $10^7$--$10^{10}$~cm$^{-3}$, CDE density in the
range $10^6$--$10^7$~cm$^{-3}$ (see Table~\ref{t:params}). The ranges of
ejected mass and outburst energy include those typical of recurrent novae
($10^{-7}< M\rs{ej} < 10^{-6}$~$M_\odot$, $10^{44} < E_0 < 10^{45}$~erg),
given the close similarities between the 2010 outburst of V407~Cyg and
those of RS~Oph and U~Sco, and the ranges typical of classical novae
($10^{-5} < M\rs{ej} < 10^{-4}$~$M_\odot$, $10^{46}< E_0 < 10^{47}$~erg;
e.g.  \citealt{2010AN....331..160B}). In the 2D simulations, the CDE
has a spherical shape (according to the symmetry of the system) and
Eq.~\ref{csm} reduces to

\begin{equation}
\rho = \rho\rs{w}\left(\frac{R\rs{cs}}{R}\right)^2+
\rho\rs{cde}~e^{[-(r/l_1)^2-(z/l_2)^2]}
\label{csm:cyl}
\end{equation}

\noindent 
with $l_1=l_2$. The CDE thickness was explored in the range 20--80~AU.
The additional 3D simulation allowed us to study the collimation of
the ejecta due to an equatorial density enhancement: in this case
we adopted the parameters of the 2D simulation that best fit the
observations and explored the effects of a disk-like CDE (according to
\citealt{2008A&A...484L...9W}) on the blast evolution, by assuming $l_1=l_2$
and $l_3<l_1$ in Eq.~\ref{csm}.

\section{Results and Discussion}
\label{s:discuss}

\subsection{Hydrodynamic evolution}

\begin{figure*}
  \centering
  \includegraphics[width=1\textwidth]{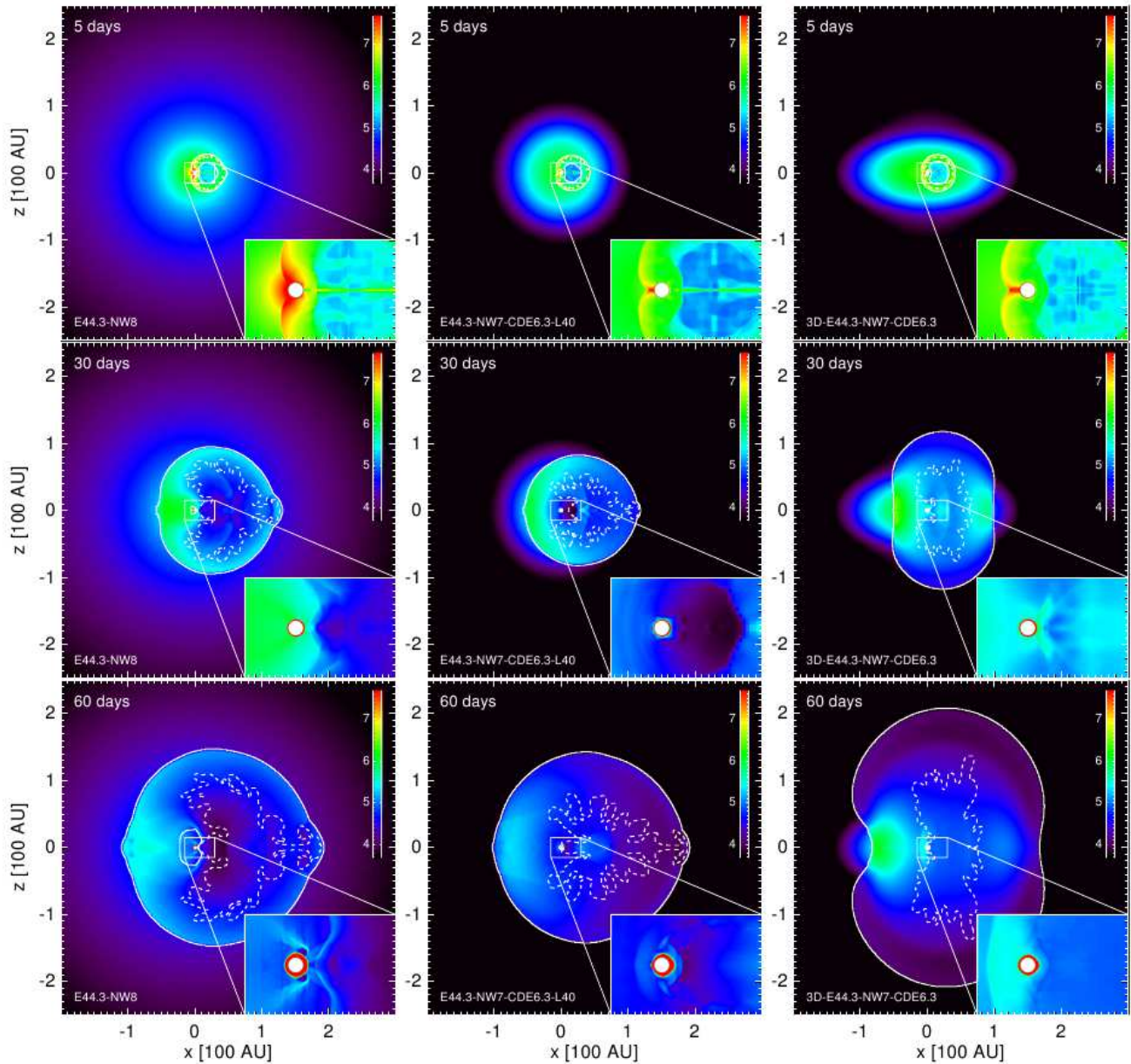}
  \caption{Cross-section images of the gas density distribution, on
    a logarithmic scale in units of cm$^{-3}$ (see the color table in
    the upper right corner of each panel), sampled at the labeled times
    for three representative V407~Cyg models. The Mira companion is at
    the origin and the WD is off-set from the origin to the right by
    15.5 AU. The different panels illustrate the evolution of the blast
    for a model without CDE (run E44.3-NW8; left panels), for a model
    with a spherical CDE (run E44.3-NW7-CDE6.3-L40; center panels),
    and for a model with a disk-like CDE (run 3D-E44.3-NW7-CDE6.3;
    right panels). Inset panels show the blast structure closer to the
    system origin. The white circle in the inset represents the Mira
    companion. The white dashed contour encloses the ejecta. The white
    solid contour denotes the regions with plasma temperature $T>10^6$~K.}
  \label{fig3}
\end{figure*}

The V407~Cyg models show a blast wave evolution that is in
some ways intermediate between those presented for RS~Oph by
\cite{2009A&A...493.1049O} and for U~Sco by \cite{2010ApJ...720L.195D}.
The WD of V407~Cyg is characterized by the massive circumstellar gas
envelope of its Mira  companion in which it sits, similar to the WD of
RS~Oph located within the dense wind of its red giant secondary, with
no evident well-defined accretion disk (unlike U~Sco).  V407~Cyg has a
much wider orbital separation than RS~Oph and U~Sco, implying that the
nova explodes initially into a relatively low density environment, as in
the case of U~Sco, despite the presence of the CDE. The size of the Mira
companion ($R\rs{cs} = 2.2$ AU) is significant compared to the size of the
system, so that the Mira partially shields the blast, as found in U~Sco.

The evolution of the blast for three representative V407~Cyg models is
illustrated in Fig.~\ref{fig3}. Post-blast gas density distributions
in the $(x,z)$ plane bisecting the system (the plane of the orbital
axis; the equatorial plane is edge on) are shown, sampled at the
labeled times. The figure also shows the distribution of ejecta
within the blast wave by highlighting regions where more than 90\%\
of the mass is material ejected in the explosion (dashed contours).
The hot post-shock plasma is dominated by thermal conduction rather
than radiative cooling, and this partially suppresses the hydrodynamic
instability that would otherwise develop during the evolution of the
blast wave (see \citealt{2005A&A...444..505O, 2008ApJ...678..274O}).
Owing to the low density environment in which the blast propagates, and
to the high post-shock temperature, there are in fact no regions within
the blast that appear dominated by radiative losses.  Thermal conduction
is expected to dominate over radiative cooling when the conduction
time-scale is shorter than the cooling time-scale or, in other words,
in plasma structures with a length-scale smaller than the Field length,
$L\rs{F} \approx 10^6 T^2/n\rs{H}$ (\citealt{1990ApJ...358..375B,
2005A&A...444..505O}), where $T$ and $n\rs{H}$ are the plasma temperature
and particle number density, respectively.  For our models,  $L\rs{F}
\approx 240$ AU, such that this length-scale is larger than the size of
the plasma structures produced in the blast.

The models all show aspherical shock morphologies, rendered by the
blast wave propagating through the inhomogeneous circumstellar gas
distribution.  In models without a CDE (hereafter NOCDE models; e.g. run
E44.3-NW8 in Fig.~\ref{fig3}), the aspherical morphology is mainly due
to the off-set of the blast wave with respect to the origin of the wind
density distribution.  A similar morphology is found in models including
a spherical CDE (hereafter SPHCDE models) with an origin coincident with
that of the wind density distribution (e.g. run E44.3-NW7-CDE6.3-L40 in
Fig.~\ref{fig3}).  Here again the morphology is determined by the off-set
of the blast wave with respect to the origin of the CDE.  At variance
with NOCDE models, however, the blast initially propagates in a medium
with a density distribution decreasing away from the companion star more
slowly than the $R^{-2}$ wind density distribution (see Fig.~\ref{fig2}).

\begin{figure}
  \centering
  \includegraphics[width=0.45\textwidth]{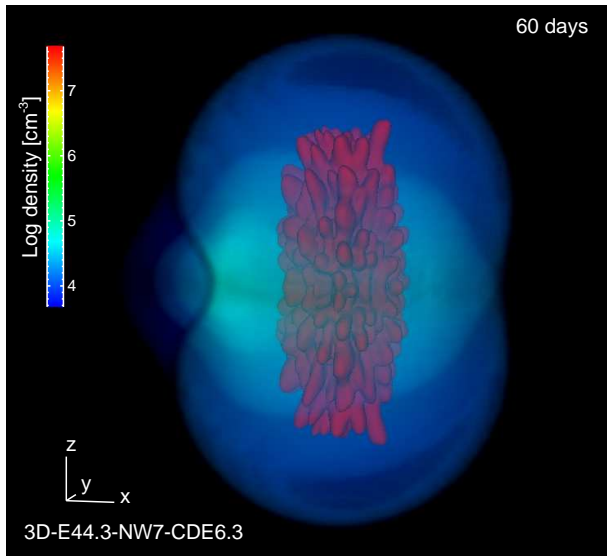}
  \caption{Three-dimensional rendering of particle number density,
    on a logarithmic scale, in units of cm$^{-3}$ (see color table in the
    upper left corner of the panel), at day 60 for the model
    3D-E44.3-NW7-CDE6.3. The ejecta are highlighted in red. The plane
    of the orbit of the central binary system lies on the $(x,y)$
    plane.}
  \label{fig4}
\end{figure}

The CDE has quite an important effect on the blast evolution if
its shape is disk-like, as in run 3D-E44.3-NW7-CDE6.3 (see right
panels in Fig.~\ref{fig3}).  In this case, the blast morphology
is similar to that found from the modeling of the RS~Oph outburst
(\citealt{2008A&A...484L...9W, 2009A&A...493.1049O}): both the blast
and ejecta are strongly collimated in polar directions, leading to a
bipolar shock morphology. It is remarkable that the shock expansion at
day 30 is almost parallel to the $z$-axis.  Figure~\ref{fig4} shows,
as an example, the collimation of blast and ejecta 60 days after the
outburst for the model 3D-E44.3-NW7-CDE6.3. A similar poleward blast
collimation was a feature of hydrodynamic simulations of the early U~Sco
blast (\citealt{2010ApJ...720L.195D}); there, the collimation was caused
by a dense accretion disk surrounding the WD.  Such a blast collimation
is also suggested by observations.  Indeed, collimation signatures
found during the 2006 RS~Oph outburst at radio, infrared, optical, and
X-ray wavelengths (\citealt{2006Natur.442..279O, 2006ApJ...652..629B,
2007A&A...464..119C, 2009ApJ...707.1168L, 2009ApJ...691..418D}) represent
a growing consensus in the literature that blast collimation is a common
feature of nova outbursts.

In all the models, the shock front propagating toward the Mira companion
is partially shielded by it and refracted around it.  As a result,
the shock converges on the rear side of the companion star, undergoing
a conical self-reflection there, and producing a wake with relatively
dense ($n\rs{H}\approx 10^6$~cm$^{-3}$) and hot ($T\approx 5\times
10^7$~K) post-shock plasma. We expect, therefore, that this structure
may lead to conspicuous observable X-ray emission.  The potential for
such a structure to explain persistent blue-ward extension of optical
emission lines was also suggested by \citet{2011A&A...527A..98S}.
At early stages, the secondary star also gives rise to a bow shock with
a temperature $T\approx 10^7$~K.  However, this feature is characterized
by relatively low densities ($n\rs{H}\approx 10^5$~cm$^{-3}$; one order
of magnitude lower than the density in the wake) and we expect only a
modest contribution to X-ray emission.

\subsection{X-ray emission}

\begin{figure*}
  \centering
  \includegraphics[width=1\textwidth]{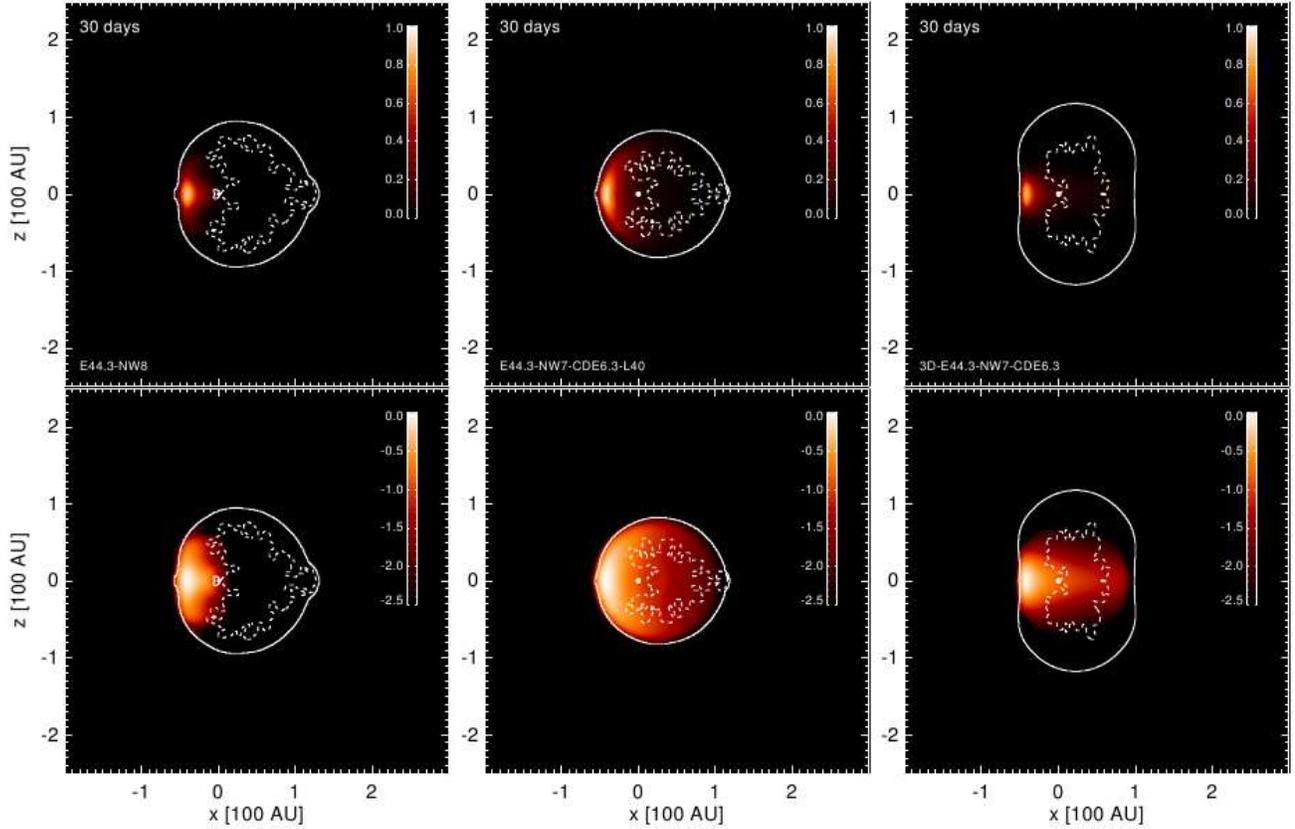}
  \caption{X-ray images (normalized to the maximum of each panel)
     in $[0.6-12.4]$ keV band projected along the line of sight of the
     blast after 30 days of evolution, corresponding to the density
     distributions in the middle panels in Fig.~\ref{fig3}. Upper panels
     are rendered with a linear surface brightness scale and lower panels
     with a logarithmic scale. Both the Mira companion and the WD lie
     on the $x$-axis. The plane of the orbit is edge on. Most of the
     X-ray emission originates from the shocked CSM. The white dashed
     contour encloses the ejecta. The white solid contour denotes the
     forward shock.}
  \label{fig5}
\end{figure*}

From the simulations, we synthesized the X-ray emission
in the $[0.6-12.4]$ keV band, using the method described by
\citet{2009A&A...493.1049O} and outlined in Sect.~\ref{s:obsanal}.
Fig.~\ref{fig5} shows the maps of X-ray emission projected along the
line of sight at day 30 (namely when a pronounced peak of X-ray
emission was observed in the 2010 outburst) for the models reported
in Fig.~\ref{fig3}. The maps are shown on linear (upper panels) and
logarithmic (lower panels) scales to highlight structures with very
different emission levels. The plane of the orbit of the central binary
system lies on the $(x,y)$ plane and is assumed to be edge on. In all
cases, most of the X-ray emission arises from the high-temperature
shocked CSM in the wake produced by the convergent shock on the rear
side of the companion star. Similar plasma structures have been
found in numerical simulations describing the outburst of RS~Oph
(e.g. \citealt{2008A&A...484L...9W}) and are generally expected as a
result of the shielding of the blast by the companion star.

\begin{figure*}
  \centering
  \includegraphics[width=1\textwidth]{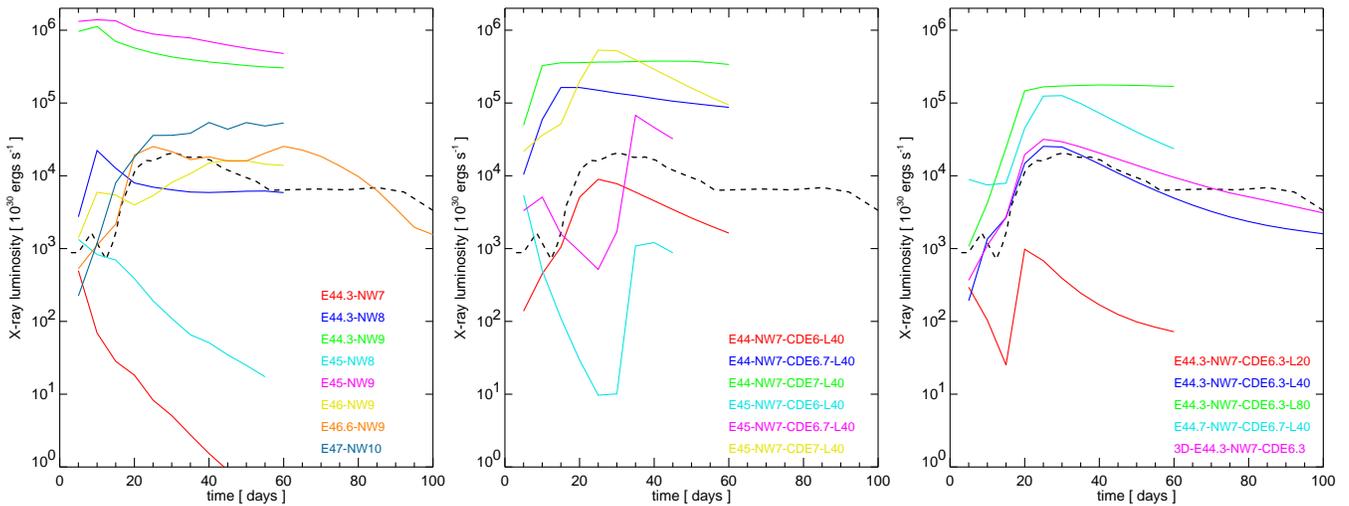}
  \caption{X-ray lightcurves in the $[0.6-12.4]$ keV band originating
     from the nova outburst (colour-coded solid lines) for the models
     reported in Table~\ref{t:params}. The left panel shows the results
     for the NOCDE models. The X-ray flux is obtained by integrating
     the synthetic X-ray emission in the whole spatial domain and takes
     into account the effects of local absorption by shocked CSM and
     ejecta. The reference black dashed line is the observed
     lightcurve reported by \citet{2011A&A...527A..98S}.}
  \label{fig6}
\end{figure*}

The X-ray lightcurves of the simulations in the $[0.6-12.4]$~keV
band are shown in Fig.~\ref{fig6}, together with the observed
lightcurve. The latter was based on the {\it Swift}/XRT count rates
reported by \citet{2011A&A...527A..98S}, considering a maximum
X-ray luminosity of $L\rs{X} = 1.2\times 10^{34}$~erg~s$^{-1}$ at
a distance of 2.7 kpc (\citealt{2011A&A...527A..98S}). As found for
U~Sco by \citet{2010ApJ...720L.195D}, in NOCDE models (left panel in
Fig.~\ref{fig6}) in general the X-ray luminosity, $L\rs{X}$,
is larger for higher explosion energy (and ejecta mass), and wind
gas density. For outburst energies and ejecta masses in the range
of values typically observed for recurrent novae (i.e. $10^{-7}<
M\rs{ej} < 10^{-6}$~$M_\odot$, $10^{44} < E_0 < 10^{45}$~erg; e.g.
\citealt{2010AN....331..160B}), the synthetic $L\rs{X}$ reaches its
maximum very quickly, within few days after the explosion, so that none
of these lightcurves fit the observed one which is characterized by a
maximum around day 30. On the other hand, assuming outburst energies and
ejecta masses typical of classical novae (i.e. in the ranges $10^{-5}
< M\rs{ej} < 10^{-4}$~$M_\odot$, $10^{46}< E_0 < 10^{47}$~erg; e.g.
\citealt{2010AN....331..160B}), the peak of X-ray emission can be
reached at later times. Among these models, that which best reproduces
the X-ray peak at day 30 is E46.6-NW9 (orange line in the left panel in
Fig.~\ref{fig6}), although it fails in reproducing the descending slope
of the observed lightcurve. This model predicts outburst energy of the
order of $\approx 5\times 10^{46}$~erg and mass of ejecta $\approx 5\times
10^{-5}$~$M_\odot$ that seem to be too high in the case of V407~Cyg that
shows many similarities with the novae RS~Oph and U~Sco.

In the presence of a CDE, the X-ray luminosity mainly depends on the
physical characteristics of the density enhancement: the greater the
size and gas density of the CDE, the larger $L\rs{X}$ (central and
right panels in Fig.~\ref{fig5}). In these models, the blast reaches
its maximum X-ray luminosity later than in NOCDE models, and the time
of maximum depends on the size and gas density of the CDE and on the
explosion energy (and ejecta mass). In particular, the time delay of
maximum X-ray luminosity is longer for lower gas density and greater size
of the CDE, and for higher explosion energy. Among the SPHCDE models,
that which best reproduces the observations is E44.3-NW7-CDE6.3-L40 (blue
line in the right panel of Fig.~\ref{fig6}): it describes a continuous
rise of X-ray luminosity up to a maximum of $L\rs{X} \approx 2\times
10^{34}$~erg~s$^{-1}$ at day 30 and then a subsequent decay phase until
day 60, with the same slope as the observed lightcurve. It is
worth noting that this model assumes outburst energies and ejecta masses
in the range of values typically observed for recurrent novae, while the
``best-fit'' NOCDE model assumes values typical of classical novae. For
E44.3-NW7-CDE6.3-L40, we extended the simulation in order to cover the
whole period in which the observed lightcurve is defined. We found that,
after day 60, the synthetic lightcurve fades faster than the observations
which seem to be characterized by a plateau between days 60 and 90. This
discrepancy could be explained as being due to the details of the density
structure of the CDE; for instance, a radial density profile of the CDE
decreasing away from the companion star slower than that modelled here
may change the descending slope of the lightcurve. Another possible
cause of the discrepancy may be some departures from the simple wind
density profile adopted here.

The effect of the shape of the CDE on the synthetic X-ray image
and lightcurve was explored by performing a 3D simulation
describing a disk-like CDE (run 3D-E44.3-NW7-CDE6.3; see
right panels in Fig.~\ref{fig3}) with the same parameters of run
E44.3-NW7-CDE6.3-L40 but with $l_1=l_2=53$ and $l_3=27$ in Eq.~\ref{csm}
(see Table~\ref{t:params}).  In this case, again most of the X-ray
emission originates in the high-temperature shocked plasma in the wake of
the companion star (see right panels in Fig.~\ref{fig5}), although the
X-ray source here is more compact than in run E44.3-NW7-CDE6.3-L40. The
X-ray lightcurve derived from run 3D-E44.3-NW7-CDE6.3 is illustrated in
the right panel of Fig.~\ref{fig6} (magenta line): the curve is similar
to that derived from run E44.3-NW7-CDE6.3-L40 and is in good agreement
with the observed lightcurve.

\section{Summary and Conclusions}
\label{s:conclusion}

We have modeled the first 60 days of evolution of the 2010 blast wave
of the nova V407~Cyg. The models explore the 3D structure of
the blast wave originating from the nova outburst, taking into account
simultaneously the radiative cooling and thermal conduction (including
heat flux saturation). Our analysis indicates the following:

\begin{enumerate}
\item the blast wave and the ejecta distribution are both aspherical
due to the inhomogeneous CSM in which the blast expands;

\item in case of a disk-like CDE, the blast and the ejecta are collimated
in polar directions, leading to a bipolar shock morphology;

\item the shock front propagating toward the Mira is partially
shielded by it, producing a wake with dense and hot post-shock plasma
on the rear side of the companion star;

\item during the evolution of the blast, most of the X-ray emission
arises from the dense and hot post-shock plasma in the wake of the
companion star;

\item the observed X-ray lightcurve can be reproduced without
including the CDE in the model only if the outburst energy and ejected
mass were of the same order of magnitude as those at the upper end
of ranges for classical novae, namely $M\rs{ej} \approx 5\times
10^{-5}~M\rs{\odot}$ and $E\rs{0} \approx 5\times 10^{46}$~erg, i.e.
values much higher than those estimated in the case of RS~Oph and U~Sco;

\item the model best reproducing the observations assumes the
presence of a CDE and suggests that the explosion energy in the 2010
outburst was $\approx 2\times 10^{44}$~erg (and the mass of ejecta
$\approx 2\times 10^{-7}~M\rs{\odot}$, i.e. values comparable to
those estimated for RS~Oph and U~Sco) and that the blast propagated
through a CDE with radius of the order of 40~AU and gas density $\approx
2\times 10^6$~cm$^{-3}$.

\end{enumerate}

\cite{2011A&A...527A..98S} note that the 2010 outburst of V407~Cyg
closely resembles the spectroscopic development of RS~Oph which is
considered the prototype of symbiotic-like recurrent novae. This fact
has suggested that V407~Cyg could belong to the same class of recurrent
novae, although its outburst was the first recorded nova event for this
system. On the other hand, as discussed by \cite{2011A&A...527A..98S}
the historical lightcurve of V407~Cyg is characterized by many gaps
and it is possible that outbursts similar to that observed in 2010 were
missed. Munari et al (1990) estimated an accretion rate of $\sim 10^{-8}
M_\odot$~yr$^{-1}$ for a WD mass of $M\rs{WD}\sim 1\,M\rs{\odot}$. If
this accreted mass were ejected in the nova event, our estimate of
the ejecta mass for the CDE case implies a recurrence time for nova
events of 20 years or so. For this accretion rate, the recurrence
timescale is consistent with higher mass white dwarfs (e.g. $M\rs{WD}\sim
1.25\,M\rs{\odot}$; e.g. \citealt{1995ApJ...445..789P}). Within the fairly
large uncertainties of accretion rate and ejecta mass estimates, this is
consistent with either a scenario in which one or more outbursts have been
missed, or with a recurrence timescale of the order of 100 years. For a
low mass WD the recurrence timescale is two orders of magnitude
longer.

In the light of the above considerations, and the results of numerical
models showing that gas does accumulate in the orbital plane of a
symbiotic binary\footnote{The gravitational accumulation of gas toward
the WD is expected to have for V407~Cyg a form similar to that for
RS~Oph. In fact, \cite{2008A&A...484L...9W} have characterized the density
enhancement by the ratio $R= V\rs{RG}/V\rs{orb}$ (where $V\rs{RG}$ is the
terminal wind velocity and $V\rs{orb}=2\pi a/P$, with $P$ the orbital
period) which for both RS~Oph and V407~Cyg is of order 1.}, we suggest
that the model best describing the 2010 outburst of V407~Cyg is that
assuming the presence of a CDE and predicting outburst energy and mass of
ejecta similar to those estimated for RS~Oph (run 3D-E44.3-NW7- CDE6.3).

Finally, the X-ray structure predicted by our model has interesting
implications for the line profiles expected to be observed during the
outburst. Simulations of the 2006 outburst of RS~Oph, that occurred
close to orbital quadrature, have shown that most of the X-ray
emission arises from a compact source propagating perpendicular to
the line of sight, which is similar to that found here.  This source
produced asymmetric and slightly blue-shifted line profiles due to
substantial X-ray absorption of red-shifted emission by shocked material
(\citealt{2009A&A...493.1049O}). These model predictions were in nice
agreement with the findings of \citet{2009ApJ...691..418D} who reported
similar asymmetric line profiles from the analysis of {\it Chandra}/HETG
observations of RS~Oph.  In the light of the above results, we expect
therefore that even in the case of V407~Cyg the line profiles should
be asymmetric and blue-shifted due to X-ray absorption of red-shifted
emission by shocked material. Interestingly, \citet{2011A&A...527A..98S}
reported asymmetric line profiles with extended blue wings during the
2010 outburst of V407~Cyg. These authors suggested that the emission
and the asymmetric line profiles could originate in the wake produced
by the convergent shock on the rear side of the companion star. We note
that the scenario proposed by \citet{2011A&A...527A..98S} is compatible
with the findings of the present paper.

\section*{Acknowledgments}
We thank an anonymous referee for useful suggestions. SO acknowledges
support from the Italian Ministero dell'Universit\'a e Ricerca (MIUR)
and from the Istituto Nazionale di Astrofisica (INAF). JJD was funded
by NASA contract NAS8-39073 to the {\it Chandra X-ray Center} (CXC)
during the course of this research and thanks the CXC director, Harvey
Tananbaum, and the CXC science team for advice and support. The software
used in this work was in part developed by the DOE-supported ASC/Alliance
Center for Astrophysical Thermonuclear Flashes at the University of
Chicago. The simulations were executed at the SCAN (Sistema di Calcolo
per l'Astrofisica Numerica) facility for high performance computing at
INAF -- Osservatorio Astronomico di Palermo.

\bibliographystyle{mn2e}
\bibliography{biblio}

\label{lastpage}

\end{document}